\documentclass[twocolumn,superscriptaddress,nobibnotes,nofootinbib,floatfix,aps,prl,citeautoscript,reprint]{revtex4-2}

\usepackage{graphicx}
\usepackage{amsmath,bm}
\usepackage{amsfonts}
\usepackage{dcolumn}
\usepackage{physics}
\usepackage[colorlinks,linkcolor=blue,citecolor=blue,urlcolor=blue]{hyperref}

\newcommand{\portland}{Department of Mechanical and Materials Engineering, Portland State University, Portland, OR 97201, USA}

\begin{document}

\title{Self-Consistent Phonon Spectral Functions and Thermal Transport \\ Beyond the Quasiparticle Approximation}

\author{Yi Xia}
\email{yimaverickxia@gmail.com; yxia@pdx.edu}
\affiliation{\portland}

\date{\today}

\begin{abstract}
Anharmonic lattice dynamics shapes phonon spectra and thermal transport, yet first-principles linewidth calculations typically assume phonon quasiparticles with sharply defined frequencies. Here, we introduce a self-consistent spectral function framework that determines interacting phonon spectral functions by dressing the internal lines of the three-phonon bubble self-energy with the spectra generated by the same interaction. The method self-consistently determines the real and imaginary parts of the self-energy on equal footing, eliminates the need for externally chosen smearing parameters to enforce energy conservation, and extends lattice dynamics and thermal transport calculations beyond the quasiparticle approximation. Applied to zincblende HgTe, self-consistency substantially broadens acoustic and optical phonon spectra, activates scattering channels inaccessible in the one-shot quasiparticle calculation, and reduces the lattice thermal conductivity by approximately a factor of five to the experimental scale while reproducing its temperature dependence, without explicitly invoking higher-order interactions. Mode-resolved spectral functions are further validated against the power spectra from direct molecular dynamics simulations. These results establish self-consistent spectral functions as a practical bridge between quasiparticle perturbation theory and fully dynamical simulations, and provide a framework readily generalizable to higher-order anharmonic processes.

\end{abstract}

\maketitle

\textit{Introduction.}
Lattice vibrations play a crucial role in phase stability~\cite{Cowley1968,SSCHA,wuttig2007phase}, heat conduction~\cite{nellis2008heat,LINDSAY2018106}, and quasiparticle dynamics~\cite{giustino2017electron, bernardi2016first} and transport~\cite{ponce2020first,lihm2025nonperturbative} in crystalline solids. Their interactions with other excitations, including electrons, spins, and excitons, are central to applications for thermoelectrics~\cite{Snyder2008,he2017advances}, photovoltaics~\cite{antonius2022theory}, and quantum devices~\cite{chirolli2008decoherence}. For predictive modeling, the key challenge is to describe not only the harmonic phonon spectrum but also the anharmonic redistribution of spectral weight arising from the underlying multi-phonon interactions~\cite{wallace1998thermodynamics,ziman1960electrons,di2026thermal}.

First-principles methods for anharmonic lattice dynamics have advanced rapidly~\cite{lindsay2016first,mcgaughey2019phonon,ravichandran2020phonon,xia2020high,mcgaughey2025phonon,luo2025tensor}. Self-consistent phonon theory~\cite{SCPH2015,tadano2022first}, temperature-dependent effective potential~\cite{TDEP2011}, and stochastic self-consistent harmonic approximation~\cite{errea2014anharmonic,bianco2017second,SSCHA} have improved the description of anharmonically renormalized phonon frequencies. Meanwhile, explicit evaluations of higher-order anharmonic scattering have shown that four-phonon~\cite{Tianli2016} and even five/six-phonon processes~\cite{xia2025first} can substantially shorten phonon lifetimes and suppress lattice thermal conductivity  ($\kappa_l$), particularly in strongly anharmonic compounds~\cite{xia2020microscopic,xia2020particlelike} or at elevated temperatures~\cite{xia2025first,Tianli2017}. In parallel, Green's function~\cite{sun2010lattice,caldarelli2022many,Isaeva:2019aa} and Wigner transport formalisms~\cite{simoncelli2019unified, Simoncelli2022} have clarified how thermal transport can be extended beyond the Peierls-Boltzmann transport equation~\cite{Peierls1929} and evaluated once interacting phonon spectral functions are known~\cite{caldarelli2022many,zeng2026lattice}. 

\begin{figure*}[htp]
    \includegraphics[width = 1.0\linewidth]{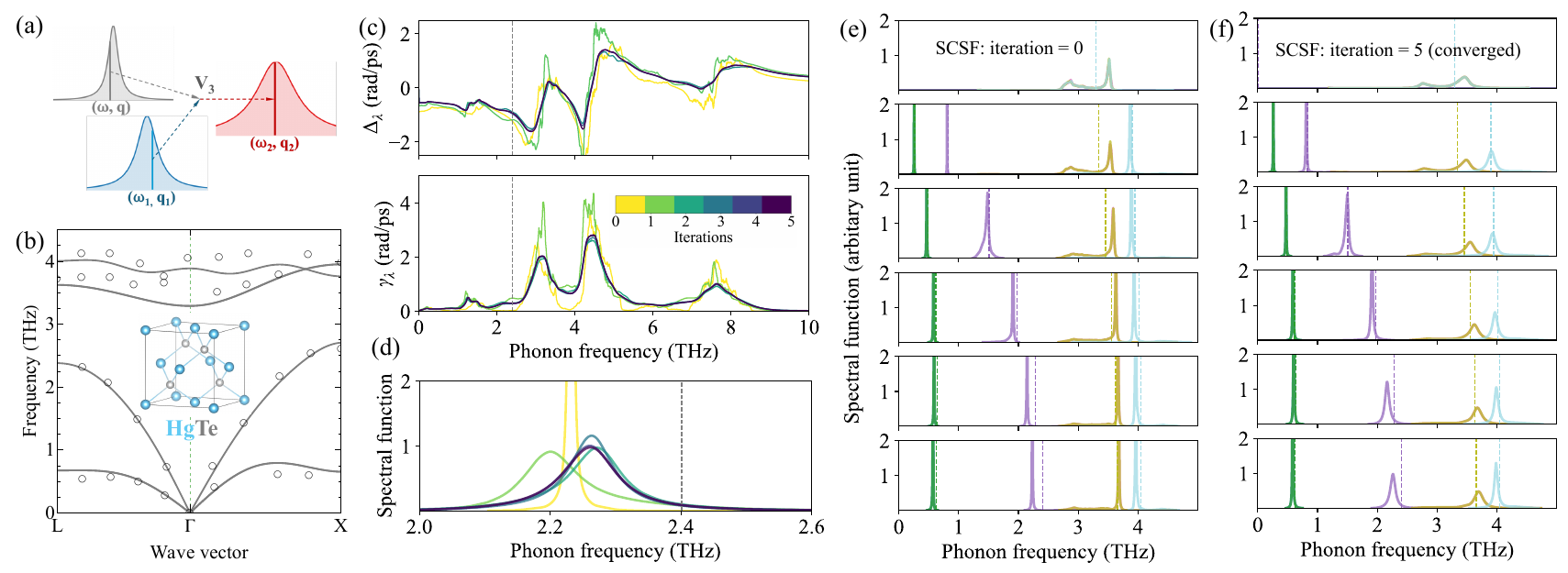}
    \caption{
    (a) Schematic illustration of three-phonon interactions in the beyond-quasiparticle picture, where interacting phonon modes are selected from spectral density distributions rather than from sharply defined quasiparticle frequencies.
    (b) Calculated harmonic phonon dispersion of zincblende HgTe in comparison with experimental measurements (open disks)~\cite{KEPA1980211,Kepa_1982}. The inset shows the zincblende crystal structure of HgTe.
    (c) Evolution of the frequency-dependent real part ($\Delta_{\lambda}$, upper panel) and imaginary part ($\gamma_{\lambda}$, lower panel) of the phonon self-energy with successive iterations of the self-consistent spectral function (SCSF) calculation for the longitudinal acoustic mode at the $L$ point of the first Brillouin zone. 
    (d) Phonon spectral function of the same mode as in (c) during the self-consistent iterations.
    (e) Phonon spectral functions at evenly sampled wave vectors along the $\Gamma$–L path, obtained from the zeroth iteration of the SCSF calculations. The corresponding fractional coordinates range from $(0,0,0)$ to $(0.5,0,0)$ in increments of 0.1.
    (f) Same as (e), but after convergence of SCSF at the 5th iteration. In (e) and (f), the two degenerate transverse acoustic modes are shown in green, the longitudinal acoustic mode in purple, the two degenerate transverse optical modes in yellow, and the longitudinal optical mode in blue. The vertical dashed lines in (c)-(f) mark the corresponding bare harmonic phonon frequencies. 2D contour plots of (e) and (f) are shown in the Supplemental Material~\cite{SM}.
    }
    \label{fig:spectra}
\end{figure*}

Despite these advances, practical calculations still treat anharmonic frequency renormalization and linewidth broadening separately: phonon frequencies may be determined self-consistently~\cite{SCPH2015,tadano2022first,errea2014anharmonic,shulumba2017lattice}, whereas linewidths are typically evaluated only once using sharply defined quasiparticle modes~\cite{shengbte,tadano2014anharmonic,han2022fourphonon, barbalinardo2020efficient,nayeb2025thermacond}, with the full frequency-dependent self-energy introduced, if at all, only in post-processing~\cite{dangic2021origin,xie2022dynamics,dangic2025lattice,xia2025lattice}. This treatment becomes inadequate when phonons are physically broadened. 
In conventional linewidth calculations, each internal phonon propagator in the corresponding Feynman diagram is assigned a single quasiparticle frequency, while energy conservation is enforced by numerically broadening the corresponding delta functions~\cite{ziman1960electrons,wallace1998thermodynamics}. Consequently, sensitivity of the calculated linewidths and $\kappa_l$ to the chosen broadening reflects not only numerical uncertainty but also an incomplete representation of the physically accessible scattering phase space~\cite{li2015ultralow,shengbte}. It is because once a phonon acquires a finite spectral width, it can participate in scattering channels that are forbidden for a mode with a single harmonic or renormalized frequency [Fig.~\ref{fig:spectra}(a)], generating a feedback between spectral broadening and scattering phase space that is absent from one-shot quasiparticle calculations.

In this Letter, we formulate and apply a self-consistent spectral function (SCSF) approach for anharmonic phonon self-energies and thermal transport. 
The central idea is to determine the interacting phonon spectral functions self-consistently by evaluating the cubic bubble self-energy with dressed internal propagators. The resulting frequency-dependent self-energy updates the spectral functions, which are then fed back into the bubble diagram until convergence.
In our test case of zincblende HgTe, this self-consistency significantly enhances phonon linewidths, yields both magnitude and temperature dependence of $\kappa_l$ in line with experiment, and produces mode-resolved spectral functions consistent with power spectra obtained from molecular dynamics simulations. Our results further show that self-consistent dressing of the cubic bubble can generate substantial additional scattering and higher-order-like temperature dependence even before explicit higher-order anharmonic diagrams are included.

\begin{figure*}[htp]
    \includegraphics[width = 1.0\linewidth]{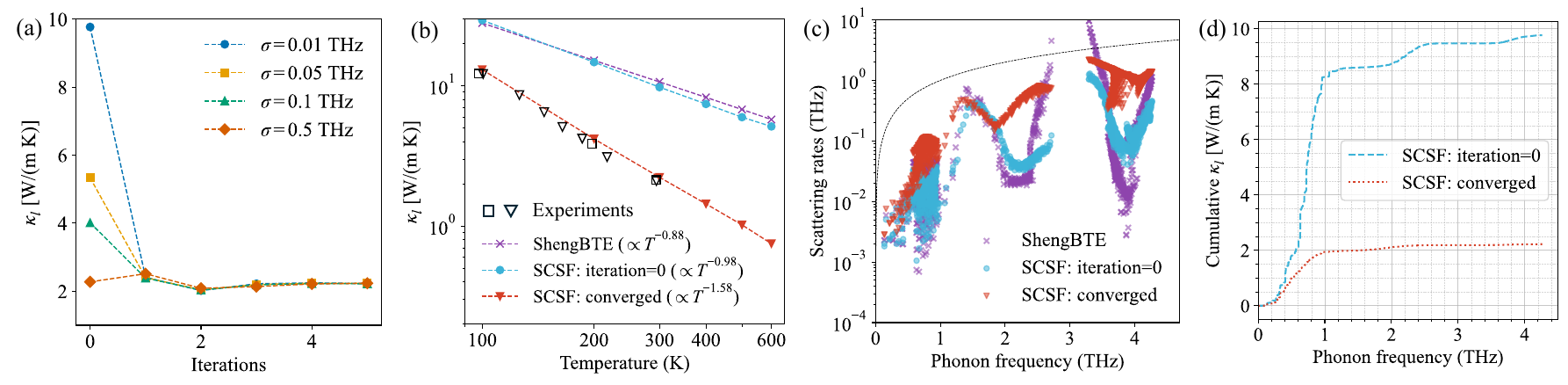}
    \caption{
    (a) Calculated lattice thermal conductivity ($\kappa_l$) of HgTe at 300~K as a function of self-consistent iterations, initialized with different Gaussian broadening parameters $\sigma$.
    (b) Temperature-dependent $\kappa_l$ obtained from the zeroth iteration and from the converged self-consistent spectral function (SCSF) calculations, compared with the conventional ShengBTE results using adaptive smearing and experimental measurements~\cite{noguera1985thermal,whitsett1972lattice}. Dashed lines indicate power-law fits to the temperature dependence of $\kappa_l$.
    (c) Mode-resolved phonon scattering rates as a function of phonon frequency at 300 K, comparing the ShengBTE results with the zeroth iteration and the converged SCSF, respectively.
    (d) Frequency-cumulative $\kappa_l$ at 300~K computed from the zeroth iteration and the converged SCSF, respectively.
    }
    \label{fig:kappa}
\end{figure*}

\textit{Theory and methods.} Considering only the lowest-order anharmonic bubble diagram~\cite{maradudin1962scattering}, the imaginary part of the frequency-dependent phonon self-energy can be expressed as the half linewidth $\gamma_{\lambda}(\omega)$ for mode $\lambda$ (composite index for wave vector and branch) under the quasiparticle approximation (QPA)
\begin{equation}
    \gamma_{\lambda}^{\rm QPA}(\omega)=
    \frac{18\pi}{\hbar^2}
    \sum_{\lambda_1,\lambda_2}
    \left|V_3^{\lambda\lambda_1\lambda_2}\right|^2
    \mathcal{F}(\omega,\omega_{\lambda_1},\omega_{\lambda_2}),
    \label{eq:gamma_qp}
\end{equation}
where $V_3$ is the cubic anharmonic vertex and momentum conservation is implicit. The phase-space factor is
\begin{equation}
\begin{split}
    \mathcal{F}(\omega,\omega_{\lambda_1},\omega_{\lambda_2})=&
    (n_{\lambda_1}+n_{\lambda_2}+1)\delta(\omega-\omega_{\lambda_1}-\omega_{\lambda_2})\\
    &+2(n_{\lambda_1}-n_{\lambda_2})\delta(\omega+\omega_{\lambda_1}-\omega_{\lambda_2}),
\end{split}
\label{eq:F_factor}
\end{equation}
with $n$ the Bose-Einstein distribution. The on-shell approximation, as commonly used in the literature~\cite{shengbte,tadano2014anharmonic,han2022fourphonon, barbalinardo2020efficient,nayeb2025thermacond}, evaluates Eq.~\eqref{eq:gamma_qp} at $\omega=\omega_{\lambda}$ and treats all participating modes as having well-defined frequencies with zero linewidth.

To go beyond this approximation, we represent each phonon by its frequency-dependent spectral function~\cite{caldarelli2022many}
\begin{equation}
    \mathcal{A}_{\lambda}(\omega) = 
    -\frac{1}{\pi}{\rm Im}~ \left[
    \frac{2\omega_{\lambda}}
    {\omega^2-\omega_{\lambda}^2-2\omega_{\lambda}\Sigma(\omega)}\right],
    \label{eq:spectral}
\end{equation}
where $\Sigma_{\lambda}(\omega)=\Delta_{\lambda}(\omega)-i\gamma_{\lambda}(\omega)$ is the phonon self-energy. Here $\Delta_{\lambda}(\omega)$ and $\gamma_{\lambda}(\omega)$ denote the real and imaginary part of $\Sigma_{\lambda}(\omega)$, respectively, and are related by the Kramers-Kronig transformation. The spectral density for finding mode $\lambda$ at frequency $\omega$ is obtained by normalizing $\mathcal{A}_{\lambda}(\omega)$ according to the convention of the adopted phonon Green's function~\cite{baggioli2019universal}: $g_{\lambda}(\omega)=\omega\mathcal{A}_{\lambda}(\omega)/\omega_{\lambda}$. This quantity can be interpreted as the mode-resolved anharmonic vibrational density of states. 
Using $g_{\lambda}(\omega)$, the many-body expression for the cubic bubble self-energy with dressed internal phonon propagators beyond the quasiparticle approximation (BQPA) can be derived
\begin{equation}
\begin{split}
    \gamma_{\lambda}^{\rm BQPA}(\omega) &  =
    \frac{18\pi}{\hbar^2}
    \sum_{\lambda_1,\lambda_2} 
     \int_{0}^{\infty}d\omega_1 
     \int_{0}^{\infty}d\omega_2
     \left|V_3^{\lambda\lambda_1\lambda_2}(\omega_1,\omega_2)\right|^2 \\
    & \times g_{\lambda_1}(\omega_1)g_{\lambda_2}(\omega_2)
    \mathcal{F}(\omega,\omega_1,\omega_2).
\end{split}
\label{eq:gamma_sc}
\end{equation}
Here, the frequency-dependent vertex follows from rewriting the conventional cubic vertex as $V_3^{\lambda\lambda_1\lambda_2}(\omega_1,\omega_2)=V_3^{\lambda\lambda_1\lambda_2}\sqrt{\omega_{\lambda_1}\omega_{\lambda2}/(\omega_1\omega_2)}$. A detailed derivation from Dyson's equation is provided in the Supplemental Material (SM)~\cite{SM}.

The self-consistency follows because the spectral functions entering the dressed bubble self-energy depend through Dyson's equation on the same complex self-energy being evaluated. We initialize $g_{\lambda}(\omega)$ from the quasiparticle results in which $\omega_1$ and $\omega_2$ are treated as sharp phonon frequencies while the self-energy is evaluated over the full frequency grid $\omega$. We then evaluate $\gamma_{\lambda}(\omega)$ from Eq.~(\ref{eq:gamma_sc}), obtain $\Delta_{\lambda}(\omega)$ through the Kramers-Kronig relation, update $\mathcal{A}_{\lambda}(\omega)$ through Eq.~(\ref{eq:spectral}), and iterate until both parts of the self-energy are converged. The resulting feedback loop from iteration step $t$, $\mathcal{A}^{(t)}\rightarrow\Sigma^{(t+1)}\rightarrow\mathcal{A}^{(t+1)}$, allows spectral broadening to modify the available scattering phase space and is absent from a one-shot spectral-function calculation. Diagrammatically, this amounts to dressing the two internal lines of the cubic bubble with propagators generated by the same interaction~\cite{mahan2000many}. It therefore extends the spirit of self-consistent phonon theory~\cite{horner1967lattice,SCPH2015} from frequency renormalization to the full complex phonon self-energy.

We note that a related Kadanoff–Baym formulation of self-consistent phonon broadening has recently been developed~\cite{di2026phonon}. While sharing the same many-body principle of dressed phonon propagation, the present SCSF formalism is derived directly from the equilibrium Dyson's equation and implemented as a real-frequency, mode-resolved first-principles scheme in which the spectral functions self-consistently renormalize the anharmonic scattering phase space. This enables direct calculations of non-Lorentzian and multi-peak phonon spectra without external smearing and permits mode-resolved validation against molecular dynamics power spectra.

We implemented the above formalism from first principles and applied it to investigate anharmonic lattice dynamics and thermal transport in zincblende HgTe. HgTe is a particularly suitable case because previous calculations based on three-phonon interaction and the Peierls-Boltzmann transport equation significantly overestimate its $\kappa_l$~\cite{ouyang2015first,rczbprx} and show pronounced sensitivity to the numerical broadening used to enforce energy conservation. Computational details, including the density functional theory setup, convergence tests, and SCSF calculations, are provided in SM~\cite{SM}.

\textit{Results and discussion.}
Fig.~\ref{fig:spectra}(b) shows the calculated harmonic phonon dispersion of HgTe, together with experimental measurements~\cite{KEPA1980211,Kepa_1982}. The overall agreement, which is consistent with the literature~\cite{ouyang2015first,rczbprx}, establishes the harmonic starting point for the spectral function analysis. We then focus on the longitudinal acoustic (LA) mode at the $L$ point in the first Brillouin zone, as a representative mode for which linewidth broadening and frequency renormalization from cubic anharmonicity are both appreciable. As shown in Fig.~\ref{fig:spectra}(c), only five iterations are required to achieve self-consistency in the frequency-dependent phonon self-energy. This self-consistent feedback smooths the sharp structures in both $\gamma_{\lambda}(\omega)$ and $\Delta_{\lambda}(\omega)$ and produces substantial changes near the bare harmonic frequency, where the spectral function is most sensitive to the self-energy. For this mode, the converged $\Delta_{\lambda}(\omega)$ becomes less negative at the bare harmonic frequency, while $\gamma_{\lambda}(\omega)$ is strongly enhanced. The corresponding spectral function in Fig.~\ref{fig:spectra}(d) reveals the physical effect of this feedback. 
Relative to the zeroth iteration, which represents a one-shot frequency-dependent spectral-function calculation without feedback of the resulting broadening into the internal propagators, the converged spectral function is much broader and shifted. Relative to the bare harmonic frequency, the spectral peak remains anharmonically softened, as commonly expected from cubic anharmonicity~\cite{Cowley1968}; however, its linewidth is no longer set by the one-shot three-phonon phase space, but is instead amplified self-consistently as broadened phonons open additional scattering channels in subsequent iterations.

Fig.~\ref{fig:spectra}(e) and (f) extend this comparison from a single mode to wave vectors along the high symmetry path $\Gamma$-$L$. The zeroth iteration spectral functions remain relatively sharp, and most peaks can still be interpreted within a quasiparticle picture. After convergence, the spectral functions are substantially broadened across both acoustic and optical branches. The effect is particularly strong for LA modes and the transverse optical (TO) modes, whose peak heights are strongly reduced. It is worth noting that near the $\Gamma$ point, the TO branch develops a non-Lorentzian lineshape with split spectral weight around the harmonic frequency. Such behavior cannot be captured by assigning a single frequency and lifetime to a phonon mode, i.e., a well-behaved quasiparticle, and it illustrates why a spectral function representation is necessary. In contrast to the pronounced changes in linewidth, self-consistency has a modest effect on the frequency shifts: both the zeroth iteration and converged spectral functions show an increasing softening of the LA branch and decreasing hardening of the TO branch from the path $\Gamma$ to $L$.

We next evaluate how the self-consistently determined spectral functions modify thermal transport. The established many-body Green's function expression~\cite{caldarelli2022many,sun2010lattice} for $\kappa_{l}$ is
\begin{equation}
\begin{split}
    \kappa_l^{\alpha\beta} = 
    \sum_{\lambda}
    \frac{\hbar^2 \omega_{\lambda}^2 v_{\lambda}^{\alpha} v_{\lambda}^{\beta}}{N_qVk_{\rm B}T^2} 
    n_{\lambda} (n_{\lambda}+1)
    \tau_{\lambda}^{\rm eff},
\end{split}
\label{eq:kappa}
\end{equation}
where $N_q$ is the number of sampled wave vectors, $V$ is the primitive cell volume, $T$ is the absolute temperature, and $v_{\lambda}^{\alpha}$ is the group velocity along $\alpha$ direction. $\tau_{\lambda}^{\rm eff}$ is the effective mode-resolved lifetime which is defined using frequency-dependent spectral functions as~\cite{xia2025lattice}
\begin{equation}
    \tau_{\mathbf q j}^{\rm eff} = 
    \pi
    \frac{
    \int_{0}^{\infty}d\omega\,
    \mathcal{A}_{\lambda}^{2}(\omega)n(\omega)[n(\omega)+1]}
    {n_{\lambda}(n_{\lambda}+1)}.
    \label{eq:tau_eff}
\end{equation}
Note that off-diagonal coherence contributions~\cite{simoncelli2019unified,Isaeva:2019aa} to $\kappa_l$ are neglected here but discussed in SM~\cite{SM} because they are small (less than 5\% at and below 300~K ) for HgTe compared with the diagonal contribution.

Fig.~\ref{fig:kappa}(a) first examines the sensitivity of $\kappa_l$ to the initial numerical broadening used in Eq.~\eqref{eq:F_factor}. At the zeroth iteration, different Gaussian widths yield noticeably different $\kappa_l$, reflecting the ambiguity introduced by external smearing of energy conservation used in Eq.~\eqref{eq:F_factor}. After self-consistency is reached, however, all calculations rapidly collapse to the same value: $\kappa_l$ converges within five iterations to 2.22~W/(m$\cdot$K) at 300~K. This insensitivity to the initialization demonstrates the robustness of our SCSF approach. 

Fig.~\ref{fig:kappa}(b) shows the impact on the temperature dependence of $\kappa_l$. 
The zeroth iteration result with 0.01~THz Gaussian broadening closely follows the conventional ShengBTE calculation with adaptive smearing~\cite{shengbte,ouyang2015first}; both exhibit nearly three-phonon-like scaling ($\propto T^{-1}$), $T^{-0.98}$ and $T^{-0.88}$, respectively, and substantially overestimate the experiments~\cite{noguera1985thermal,whitsett1972lattice}. Such an agreement reveals that the beyond-quasiparticle effects are weak in the absence of self-consistent spectral functions.
In contrast, the converged SCSF strongly suppresses $\kappa_l$ and yields a steeper $T^{-1.58}$ dependence, bringing both the magnitude and temperature dependence much closer to the experiments. 
This faster-than-$T^{-1}$ decay, often associated with higher-order anharmonic scattering~\cite{Tianli2016}, already emerges when the anharmonic Hamiltonian is restricted to cubic order but the bubble self-energy is treated self-consistently. Importantly, the comparison between the zeroth iteration and converged SCSF uses identical cubic force constants and therefore isolates the additional scattering generated by self-consistent spectral broadening. However, we note that this does not imply that explicit quartic anharmonicity is negligible in HgTe.

To identify the microscopic origin of this suppression of $\kappa_{l}$, Fig.~\ref{fig:kappa}(c) compares mode-resolved scattering rates at 300 K. The zeroth iteration rates largely resemble the results from ShengBTE, with differences traceable to the use of fixed versus adaptive broadening. After self-consistency is reached, the scattering rates increase broadly across the spectrum, with enhancements exceeding one order of magnitude for some modes. The enhancement is especially pronounced for the high-frequency transverse acoustic (TA) modes, LA modes near the Brillouin zone boundary, and almost all optical modes. This is consistent with the spectral functions in Fig.~\ref{fig:spectra}, which show that these modes acquire large intrinsic linewidths. The frequency-cumulative $\kappa_l$ in Fig.~\ref{fig:kappa}(d) further shows that the largest reduction in $\kappa_{l}$ comes from modes below approximately 1 THz, mainly TA modes, while additional reductions arise from enhanced scattering of higher-lying acoustic and optical modes. 

\begin{figure}[htp]
    \includegraphics[width = 0.75\linewidth]{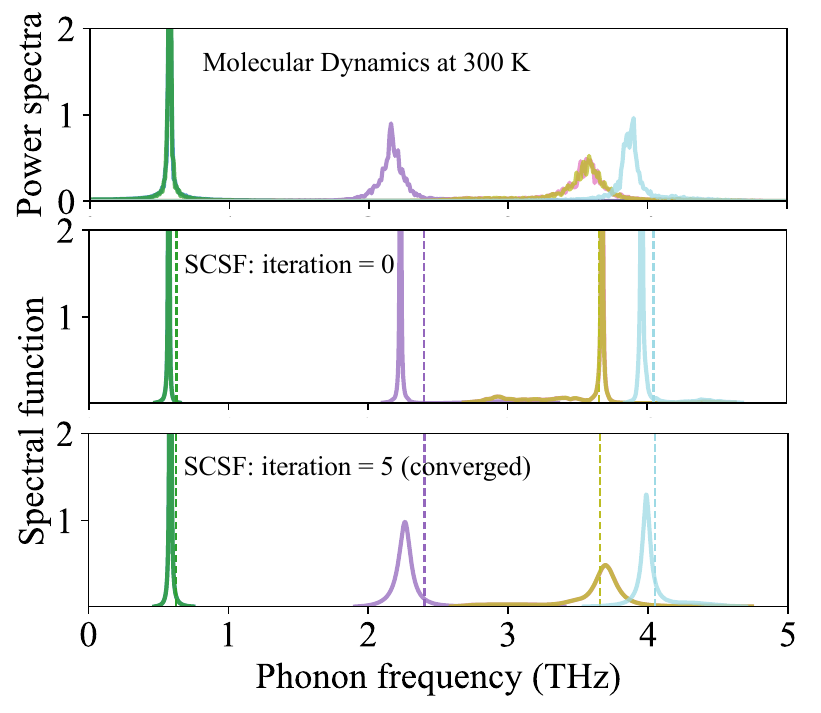}
    \caption{
    Comparison of the phonon spectra for all six modes at the $L$ point of HgTe at 300 K. The upper panel shows the power spectra obtained from molecular dynamics simulations. The middle and lower panels show the phonon spectral functions obtained from the zeroth iteration and from the converged SCSF, respectively. Vertical dashed lines indicate the bare harmonic phonon frequencies.
    }
    \label{fig:md}
\end{figure}

\begin{figure*}[htp]
    \includegraphics[width = 1.0\linewidth]{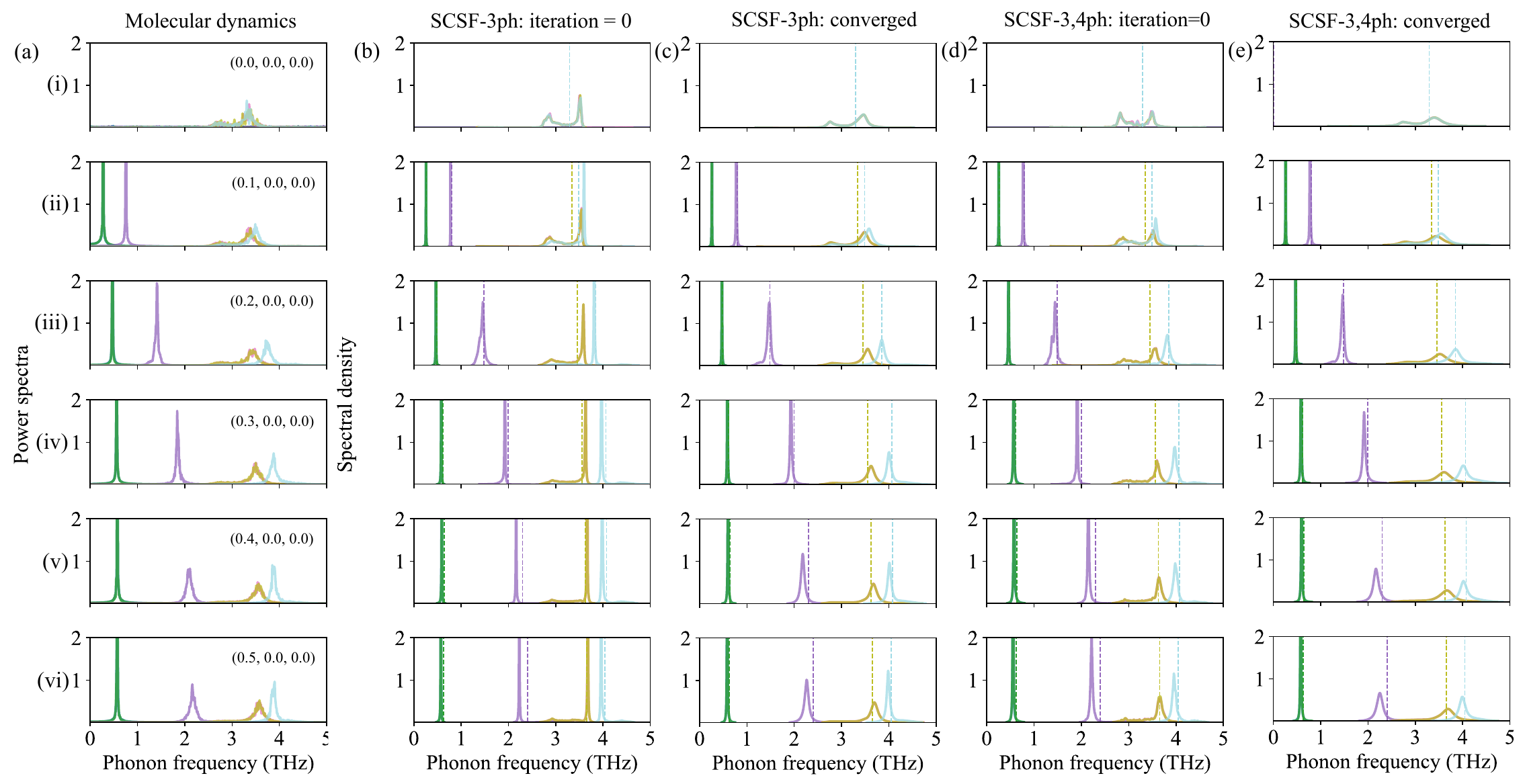}
    \caption{
    Comparison of the full frequency-dependent phonon spectral functions of HgTe at 300~K at the fractional phonon wave vectors indicated in panel (a), obtained from (a) classical molecular dynamics; (b) zeroth-iteration SCSF with the cubic bubble diagram, corresponding to the perturbative full frequency-dependent self-energy; (c) converged SCSF with the cubic bubble diagram; (d) zeroth-iteration SCSF with both the cubic bubble and quartic sunset diagrams; and (e) converged SCSF with both diagrams. Rows (i)--(vi) correspond to the six fractional phonon wave vectors shown in panel (a).
    }
    \label{fig:em_allsf}
\end{figure*}

Agreement with experimental $\kappa_l$ is encouraging but not by itself decisive, because transport coefficients can benefit from error cancellation~\cite{pbte2018}. A more direct validation of our approach is to compare the calculated spectral functions with mode-resolved power spectra obtained from molecular dynamics (MD) simulation. We therefore trained a machine learning interatomic potential using the moment tensor potential approach~\cite{shapeev2016moment}, with training structures generated from on-the-fly learning~\cite{jinnouchi2019fly} within VASP~\cite{Vasp1,Vasp2,Vasp3}. Mode-resolved power spectra were computed from MD trajectories using the DynaPhoPy~\cite{carreras2017dynaphopy} approach (see computational details in SM~\cite{SM}). Fig.~\ref{fig:md} compares the MD power spectra with the SCSF for all six modes at the $L$ point at 300 K. We see that the zeroth iteration spectral functions are rather narrow, while the converged SCSF closely matches the MD power spectra after applying the appropriate temperature normalization factor to the latter. The agreement is achieved at the mode level, not only at the level of a Brillouin-zone-averaged transport coefficient. Similar trends for phonon wave vectors along the full $\Gamma$-$L$ path are shown in Fig.~\ref{fig:em_allsf} in End Matter. This comparison supports the interpretation that our SCSF approach captures rich anharmonic broadening physics that is missing from the one-shot quasiparticle linewidth calculation.

Recognizing the potential importance of four-phonon scattering in HgTe~\cite{rczbprx,luo2025tensor,wu2026origin}, we further examined quartic anharmonicity by self-consistently including the quartic sunset diagram. As shown in Appendix~A of the End Matter, self-consistency remains essential when this contribution is included. However, a simple additive treatment of cubic and quartic self-energies tends to overbroaden several optical modes relative to MD, indicating the potential importance of mixed-vertex interactions that couple cubic and quartic anharmonicity. An example calculation and further discussion are provided in Appendix~B of the End Matter.

We note that our SCSF framework is not limited to HgTe, but should be broadly applicable to both weakly and strongly anharmonic materials. As a complementary benchmark in the weakly anharmonic regime, we also applied SCSF to Si. The converged $\kappa_{l}$ remains close to the conventional ShengBTE result, demonstrating that the self-consistent treatment smoothly approaches the quasiparticle limit when intrinsic spectral broadening is small (see SM~\cite{SM}). More generally, for modest anharmonicity, finite spectral broadening can activate scattering channels that are inaccessible to sharply defined harmonic phonons, for example in compounds with large acoustic-optical gaps and correspondingly restricted three-phonon phase space~\cite{Lindsay2013, yang2019stronger, rczbprx}. In strongly anharmonic systems, where phonon linewidths are intrinsically large, a spectral-function treatment can become even more important. Materials characterized by substantial higher-order scattering despite restricted lower-order scattering, including AgCrSe$2$~\cite{xie2020first}, cuprous halides (CuCl~\cite{mukhopadhyay2017curious,kundu2025revisiting}, CuBr, and CuI), silver halides (AgCl, AgBr, and AgI)~\cite{ouyang2023role}, as well as Cu$_2$O, and SnSe~\cite{aseginolaza2019phonon}, therefore provide promising systems for further applications. Our preliminary calculations for CuCl show that the SCSF approach reduces $\kappa_l$ by more than a factor of five and brings it close to experiment, further supporting the generality of the mechanism (see Fig.~S7 and relevant discussion in SM~\cite{SM}).

It is also instructive to note the parallel development in electronic transport, where recent approaches have progressed from self-consistent quasiparticle descriptions~\cite{holstein1964theory,lihm2024self} toward treatments retaining the full frequency dependence of the electronic Green’s function and spectral function~\cite{lihm2025nonperturbative,lihm2026beyond}. Together with the present phonon formulation, these advances point toward a unified self-consistent beyond-quasiparticle framework for coupled electron–phonon dynamics and transport, in which both electronic and vibrational excitations are represented by their full interacting spectral functions.

\textit{Conclusion.} In summary, we have developed a self-consistent spectral function framework for anharmonic phonons and lattice thermal transport beyond the quasiparticle approximation. 
By iteratively dressing the internal phonon lines of the cubic bubble diagram with the spectral functions generated by the same self-energy, the approach determines interacting phonon spectra self-consistently and captures the feedback between spectral broadening and scattering phase space. These converged spectra can be directly incorporated into established Green's function transport formulations beyond the quasiparticle approximation.
Applied to zincblende HgTe, this feedback produces strongly broadened phonon spectra and substantially suppresses $\kappa_l$, bringing both its magnitude and temperature dependence close to experiment; the resulting spectral functions are further validated against mode-resolved power spectra derived from molecular dynamics simulations. We expect this approach to be broadly applicable, particularly to materials with restricted three-phonon scattering phase space or strong anharmonicity. More broadly, the framework, demonstrated here for cubic and cubic-plus-quartic self-energies, provides a bridge between quasiparticle perturbation theory and fully dynamical simulations and can be systematically extended to still higher-order and mixed-vertex anharmonic processes.

\begin{acknowledgments}
\textbf{Acknowledgments:} Y. X. acknowledges support from the U.S. National Science Foundation through award CBET-2445361, support from the Faculty Development Program at Portland State University, and computing resources provided by Bridges2 at Pittsburgh Supercomputing Center through ACCESS allocations mat220006p and mat220008p. ACCESS is supported by National Science Foundation grants 2138259, 2138286, 2138307, 2137603, and 2138296. Y. X. is grateful to Z. J. W. for encouragement and support during the preparation of this manuscript.
\end{acknowledgments}

\textit{Note added.} 
While preparing this manuscript, we became aware of two related studies~\cite{castellano2025fluctuation,di2026phonon}. Ref.~\cite{castellano2025fluctuation} formulates related physics using the fluctuation-dissipation theorem, while Ref.~\cite{di2026phonon} develops a general Kadanoff–Baym treatment of phonon collisional broadening and heat transport. Here, we derive the SCSF equation directly from the equilibrium Dyson's equation by evaluating the cubic bubble self-energy with dressed internal propagators through their spectral representations, and further extend the self-consistent framework to include the contribution from the quartic sunset diagram. The resulting real-frequency formulation provides a practical first-principles route to self-consistent non-Lorentzian phonon spectra without external smearing, which we further validate against mode-resolved molecular dynamics power spectra.

\bibliography{CuSbS}



\clearpage

\twocolumngrid
\begin{center}
  \textbf{\large End Matter}
\end{center}

\textbf{Appendix A: Impact of quartic anharmonicity}

To investigate the effects of higher-order anharmonicity on the computed phonon spectral functions, we further consider the quartic sunset diagram~\cite{xiao2023anharmonic,xia2025lattice}, commonly referred to as four-phonon (4ph) interactions~\cite{Tianli2016}, in addition to the cubic bubble diagram describing three-phonon (3ph) interactions. Specifically, we generalize Eq.~(4) of the main text by incorporating the sunset diagram following our previous derivation~\cite{xia2025lattice}:
\begin{widetext}
\begin{equation}
\begin{split}
    \gamma_{\lambda}^{\rm BQPA-4ph}(\omega)  = &
    \frac{96\pi}{\hbar^2}
    \sum_{\lambda_1,\lambda_2,\lambda_3} 
     \int_{0}^{\infty}d\omega_1 
     \int_{0}^{\infty}d\omega_2
     \int_{0}^{\infty}d\omega_3~
     \left|V_4^{\lambda\lambda_1\lambda_2\lambda_3}(\omega_1,\omega_2,\omega_3)\right|^2 \\
     & \times g_{\lambda_1}(\omega_1)g_{\lambda_2}(\omega_2)g_{\lambda_3}(\omega_3)
    \mathcal{F}^{\rm 4ph}(\omega,\omega_1,\omega_2,\omega_3),
\label{eq:gamma_4ph}
\end{split}
\end{equation}
\begin{equation}
\begin{split}
{\rm with~} \mathcal{F}^{\rm 4ph}(\omega,\omega_1,\omega_2,\omega_3) & =  \left[(n_1+1)(n_2+1)(n_3+1)-n_1n_2n_3\right] \left[\delta(\omega+\omega_1+\omega_2+\omega_3)-\delta(\omega-\omega_1-\omega_2-\omega_3)\right] \\
& +3\left[n_1(n_2+1)(n_3+1)-(n1+1)n_2n_3\right] \left[\delta(\omega-\omega_1+\omega_2+\omega_3)-\delta(\omega+\omega_1-\omega_2-\omega_3)\right].
\end{split}
\end{equation}
\end{widetext}

The resulting full frequency-dependent imaginary part of the self-energy is $\gamma_{\lambda}^{\rm BQPA}=\gamma_{\lambda}^{\rm BQPA-3ph}+\gamma_{\lambda}^{\rm BQPA-4ph}$, which enters our self-consistent spectral function (SCSF) formalism. Owing to the substantial computational cost of evaluating $\gamma_{\lambda}^{\rm BQPA-4ph}$ on a dense frequency grid while simultaneously achieving self-consistency, the SCSF calculations including both the bubble and sunset diagrams were performed for HgTe at 300~K using a relatively sparse $10\times10\times10$ phonon wave-vector mesh.

Figure~\ref{fig:em_allsf} compares the full frequency-dependent phonon spectral functions of HgTe obtained from classical molecular dynamics (MD) simulations with those from SCSF calculations including either only the cubic bubble diagram or both the cubic bubble and quartic sunset diagrams. Without self-consistency, substantial discrepancies from the MD results are observed for both the zeroth-iteration SCSF-3ph and SCSF-3,4ph calculations. In particular, the zeroth-iteration SCSF-3ph calculation substantially underestimates the spectral broadening [Fig.~\ref{fig:em_allsf}(a) vs. Fig.~\ref{fig:em_allsf}(b)], whereas the zeroth-iteration SCSF-3,4ph calculation produces an incorrect redistribution of spectral weight [Fig.~\ref{fig:em_allsf}(a) vs. Fig.~\ref{fig:em_allsf}(d)], most noticeably for the zone-center optical modes. These results demonstrate that simply evaluating the corresponding frequency-dependent self-energies perturbatively is insufficient to reproduce the anharmonic spectral features.

Upon convergence, the SCSF-3,4ph spectral functions become qualitatively comparable to those obtained from the converged SCSF-3ph calculation [Fig.~\ref{fig:em_allsf}(c) vs. Fig.~\ref{fig:em_allsf}(e)]. A closer comparison with the MD spectra, however, reveals systematic differences. Including the quartic sunset diagram generally increases the phonon broadening. This additional broadening appears to improve the agreement with MD for some acoustic modes, whereas it tends to overbroaden most optical modes and suppress their spectral peaks relative to the MD results. These results confirm that quartic anharmonicity can further modify the spectra, but also show that the discrepancy cannot be understood simply as a missing additive 4ph scattering rate. The enhanced broadening consequently reduces the calculated $\kappa_{l}$ to $1.31$~W/(m,K) at 300~K, which is lower than the experimental value discussed in the main text.

We note that this reduced $\kappa_{l}$ should nevertheless be interpreted with caution. First, the relatively sparse $10\times10\times10$ wave-vector mesh used in the SCSF-3,4ph calculations may not be sufficiently converged for a quantitative evaluation of $\kappa_{l}$. More importantly, simply adding the quartic sunset self-energy to the cubic bubble self-energy neglects coupling between cubic and quartic anharmonic interactions. Such mixed diagrams can make negative contributions to the linewidth; one representative diagram is quantified explicitly in the following section. In addition, when quartic anharmonicity is treated explicitly in the phonon dynamics, the corresponding higher-order contributions to the heat current operator should, in principle, also be included in the calculation of $\kappa_{l}$~\cite{di2026thermal}; such contributions have been shown to be significant in previous work~\cite{sun2010lattice}.

Despite these limitations, Fig.~4 clearly demonstrates that for both cubic-only and cubic-plus-quartic treatments, the zeroth-iteration spectra differ substantially from the converged results and the MD reference, highlighting the need for a self-consistent treatment of the frequency-dependent phonon self-energy.

\textbf{Appendix B: Mixed-vertex interactions}

\begin{figure}[htp]
    \includegraphics[width = 0.75\linewidth]{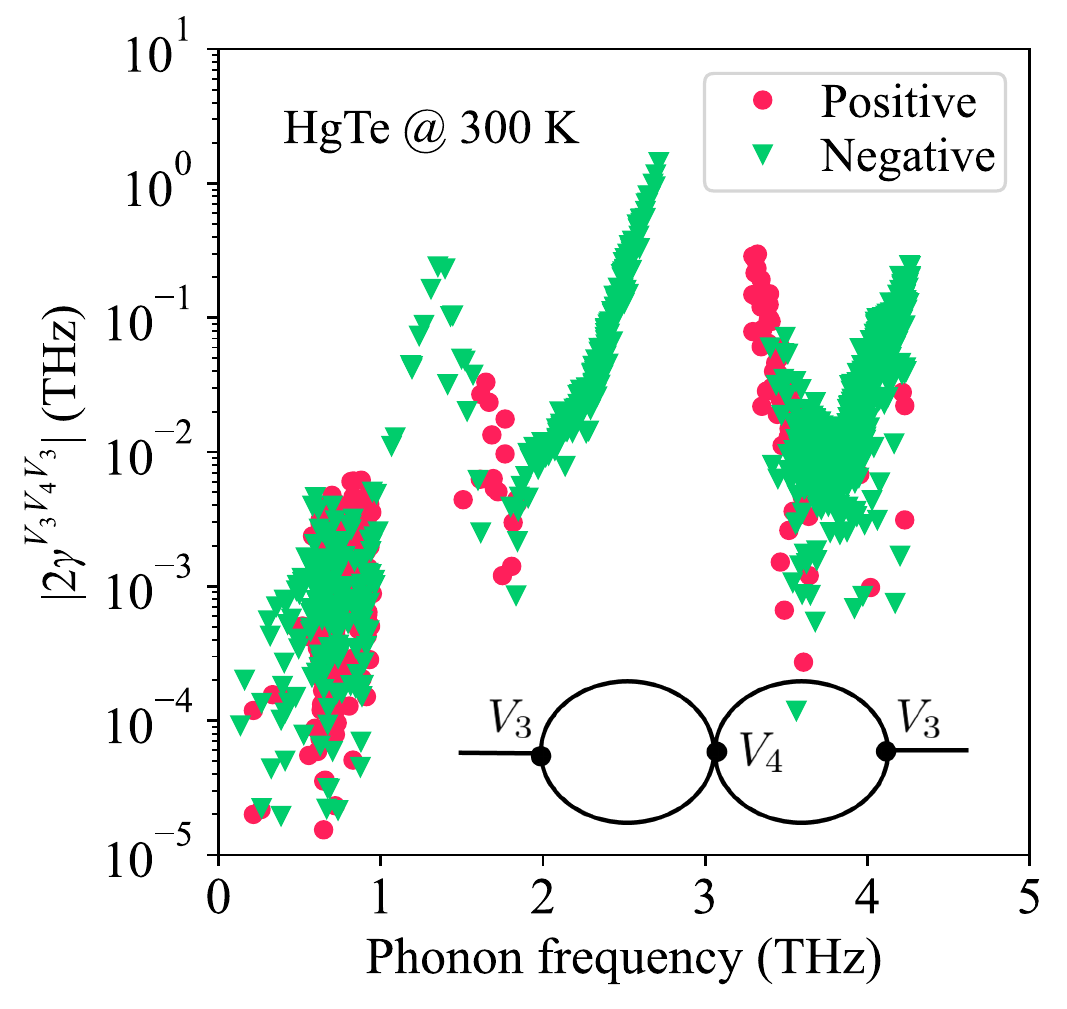}
    \caption{
    Computed mixed-vertex $V_3V_4V_3$ linewidth for HgTe at 300~K, arising from the Feynman diagram in the inset. Positive and negative linewidth values are indicated by red circles and green triangles, respectively.
    }
    \label{fig:em_mixed}
\end{figure}

Within the conventional framework of anharmonic phonon scattering, the conventional 3ph and 4ph interactions arise from double-vertex diagrams, with analogous extensions to higher-order five- and six-phonon processes~\cite{xia2025first}. A rigorous diagrammatic treatment, however, also contains mixed-vertex diagrams involving different orders of anharmonicity~\cite{tripathi1974self,di2026thermal}, such as coupled cubic and quartic interactions. To the best of our knowledge, the quantitative importance of these mixed-vertex contributions has not yet been systematically examined from first principles, although a pioneering study using a simplified model composed of a diatomic linear lattice has shown that they can yield negative contributions to phonon linewidths~\cite{bhandari1992phonon}. We note that such negative contributions are not unphysical. It is because while the total phonon linewidth obtained after summing all relevant diagrams must remain nonnegative, individual diagrammatic terms can have either sign. In contrast, conventional double-vertex contributions are positive definite because they involve an anharmonic vertex multiplied by its complex conjugate.

To illustrate the potential importance of mixed-vertex interactions, we evaluate the mixed cubic-quartic contribution associated with a representative $V_3V_4V_3$ diagram shown in the inset in Fig.~\ref{fig:em_mixed} (see theoretical derivation and comparison with double-vertex diagrams in SM~\cite{SM}) and find both positive and negative linewidth contributions across a broad phonon-frequency range. 
The sign variation demonstrates that mixed anharmonic contributions can, in principle, either enhance or partially cancel the conventional broadening. Although the single $V_3V_4V_3$ diagram evaluated here does not constitute a complete treatment, it shows why simply adding positive 3ph and 4ph linewidths is not systematically complete. These results therefore highlight the need to include a comprehensive set of mixed-vertex diagrams when accounting for quartic anharmonicity, together with their interplay with self-consistent phonon spectral functions, for a more complete treatment of higher-order anharmonicity in future work.

\end{document}